\documentclass[twocolumn,prl,nobibnotes,altaffilletter,amsmath,amssymb,amsfonts]{revtex4}
\usepackage[latin1]{inputenc}
\usepackage{graphicx}
\usepackage{amsmath}
\usepackage{latexsym}
\usepackage{epsfig}
\usepackage{bm}	
\usepackage{enumerate}
\usepackage{lipsum}

\usepackage[dvipsnames]{color}

\begin{document}
\title{Multidimensional Stationary Probability Distribution for Interacting Active Particles}

\author{Claudio Maggi$^1$}
\email{claudio.maggi@roma1.infn.it}
\author{Umberto Marini Bettolo Marconi$^2$}
\author{Nicoletta Gnan$^3$}
\author{Roberto Di Leonardo$^{1,4}$}

\affiliation{$^1$Dipartimento di Fisica, Universit\`a di Roma ``Sapienza'', 
I-00185, Roma, Italy }
\affiliation{$^2$Scuola di Scienze e Tecnologie, Universit\`a di Camerino, Via Madonna delle Carceri, 62032, Camerino, INFN Perugia, Italy}
\affiliation{$^3$CNR-ISC, UOS Sapienza, P.le A. Moro2, I-00185, Roma, Italy}
\affiliation{$^4$CNR-IMIP, UOS Roma, Dipartimento di Fisica Universit\`a Sapienza, I-00185 Roma, Italy}

\date{\today}
\begin{abstract}
We derive the stationary probability distribution for a non-equilibrium system composed by an arbitrary number of degrees of freedom that are subject to Gaussian colored noise and a conservative potential. This is based on a multidimensional version of the Unified Colored Noise Approximation. By comparing theory with numerical simulations we demonstrate that the theoretical probability density quantitatively describes the accumulation of active particles around repulsive obstacles. In particular, for two particles with repulsive interactions, the probability of close contact decreases when one of the two particle is pinned. Moreover, in the case of isotropic confining potentials, the radial density profile shows a non trivial scaling with radius.  Finally we show that the theory well approximates the ``pressure" generated by the active particles allowing to derive an equation of state for a system of non-interacting colored noise-driven particles.  

\end{abstract}

\maketitle

\section*{Introduction}

A generic system, that is in thermal equilibrium at a temperature $T$, will be found in the neighbourhood of a configuration of energy $E$ with a probability density given by the  the Boltzmann factor $\exp[-E/k_B T]$ \cite{StatLect,StatMechBooks}.
As an example, two Brownian colloidal particles, interacting through a conservative attractive force, will show an increased probability density for the low energy bound state. A quite different behaviour is observed in active particle systems~\cite{ReviewsActive}.
Generally speaking, active matter is composed by biological or synthetic objects that are capable of absorbing energy from the environment and convert it into different kinds of persistent motions. Even when stationary states are reached, the probability distributions can display large deviations from their equilibrium Boltzmann counterparts. Those deviations are not just a matter of quantity, but a radically different qualitative behaviour may be observed, like the widespread tendency to accumulate around repulsive objects. This ``attraction for repulsion'' is responsible for phenomena like, particle accumulation at solid walls~\cite{Walls2} or the formation of bound states between repulsive objects~\cite{ActiveDepletion,Ivo}. Our intuitive notion that particles like to stay where external forces attract them to is biased by our familiarity with equilibrium statistical mechanics and needs to be replaced by novel statistical mechanics concepts that are capable to describe the stationary probability distributions  in systems of interacting active particles.
In this context some schematic models have been proposed to model the dynamics of active particles as, for example, the ``run and tumble" (RnT) model. The RnT dynamics is appropriate to describe the motion of bacteria such as \textit{E. coli}~\cite{Berg,Schnitzer,CatesEPL,CatesPRL} that swim along almost straight runs interrupted by random reorientations. In the case of colloids propelled by chemical reactions the ``active Brownian" model describes the motion of particles pushed by a force of constant magnitude that gradually reorient by rotational Brownian motion~\cite{Janus1,Janus2,Janus3,Bocquet}.
However, despite the simplicity of the dynamics of these systems, it is hardly possible to find the {analogue} of the Boltzmann distribution. Indeed, the Boltzmann {prescription} assigns a precise {weight} to a given configuration of positions and momenta of particles at equilibrium. These particles are {embedded}  in a space of dimensionality $d$, {are subject} to arbitrary external fields and mutually interact via whatsoever potential~\cite{StatLect,StatMechBooks}.
{On the contrary, in the case of active particles  the  exact stationary probability distribution is known only in rare instances} as, for example, in the 1-dimensional RnT model in an external force field~\cite{CatesEPL}. The impossibility of writing explicitly the stationary probability density prevents one from applying the standard methods of statistical mechanics. 
A Gaussian colored-noise model (GCN) can be used to reproduce the dynamics of passive colloidal particles immersed in a bath of dense swimming bacteria~\cite{StructSim,Capillars}. This model has been intensively studied in the past as the simplest model that could elucidate the basic physics of systems subject to time-correlated noise. Interestingly GCN was originally used to interpret the behaviour of very different physical systems such as noisy electronic circuits~\cite{Circuits} and dye-laser radiation~\cite{Dye}. The analytical study of GCN-driven systems resulted very challenging and led to the development of different approximation schemes aimed to reduce the complexity of the GCN-model to a tractable level~\cite{HanggiRev,HanggiUcna}. Among these approaches one emerges by having a number of advantages with respect to the others. This is the \emph{Unified Colored Noise Approximation} (UCNA) developed by H\"{a}nggi and Jung~\cite{HanggiUcna} that, under certain conditions, describes both the small and the large correlation-time regimes, both in the high and low-friction limit~\cite{Inertia}. More importantly for the present work the UCNA scheme can be generalized to a phase space of {arbitrary} dimensionality~\cite{UcnaD}.
In this work we report, for the first time, the explicit formula of the stationary probability 
(obtained within the UCNA) for a system that is subject to a generic conservative potential and that is composed by an arbitrary number of degrees of freedom. We name this ``multidimensional unified colored noise approximated stationary probability" {(MUCNASP)}. The MUCNASP plays basically the same role {as}  the Boltzmann distribution for the approximated GCN-driven system. We show how the MUCNASP allows to predict several non-equilibrium properties of the active system in experimentally relevant cases where a simple external potential acts on a small number of degrees of freedom.
We focus on the case of steep repulsive interactions and spherically symmetric external potentials. In all these situations we use numerical simulation to test the quality of the approximation {and  find that the GCN-driven 
and the RnT particles display  a strikingly similar behavior}. 
In particular we show how our approximated probability density captures very well the accumulation of the active particles around repulsive obstacles. 
Moreover the theory describes well the dependence on dimensionality of the probability density function when the active particles are confined by a circular repulsive wall.
Understanding how the concept of pressure generalizes to active matter has become recently the subject of intense theoretical research~\cite{Brady,CatesPressure,CatesPressure2}.
In this context we show how our theoretical probability density allows us to derive the pressure that the active particles exert on the repulsive walls and leads to the derivation of equations of state for the non-interacting active particle system. 
Finally we discuss the most relevant limitations of the present theory and  suggest new routes to follow in the theoretical study of active matter.

\section*{Results}

We consider the {following} set of stochastic differential equations:

\begin{equation}
\label{eq:sys2}
\dot{\textbf{x}} 
= -\nabla 
\Phi+
\bm{\eta}
\end{equation}

\noindent where the {position} variables $\mathbf{x}=(x_1,...,x_N)$ 
are determined by the deterministic velocity $-\nabla \Phi$ generated by a 
conservative potential $\Phi(\mathbf{x})$ and by a set of stochastic processes
$\bm{\eta}=(\eta_1,...,\eta_N)$. We assume that these are $N$ independent 
Gaussian processes with zero mean $\langle \eta_j \rangle=0$ and exponential time-correlation: 
$\langle \eta_i(t) \eta_j(s) \rangle = \delta_{ij} \frac{D}{\tau} e^{-|t-s|/\tau}$.
Here $D$ is the diffusion coefficient characterizing the amplitude of the noise and 
$\tau$ is its relaxation time. Note that here we absorb the mobility $\mu$ in 
the velocity field $-\nabla \Phi=\mu \mathbf{f}$, where $\mathbf{f}=-\nabla U $ 
is the deterministic force generated by the potential energy function $U(\mathbf{x})= \Phi(\mathbf{x})/\mu $.
By using the UCNA Eq.~(\ref{eq:sys2}) 
reduces to the (Stratonovich) Langevin equation~\cite{UcnaD}:

\begin{equation} \label{eq:ucnad}
 [\mathbb{I}+\tau \mathbb{H}(\mathbf{x})] \, \dot{\mathbf{x}} = 
 -\nabla \Phi(\mathbf{x})+D^{1/2}\bm{\Gamma}
\end{equation}

\noindent where $\mathbb{I}=\delta_{ij}$ is the identity matrix, $\mathbb{H}(\mathbf{x})~=~\partial_{x_i}\partial_{x_j} \Phi(\mathbf{x})$ is the Hessian associated with the potential $\Phi$, and  $\bm{\Gamma}$ is a set of independent
white-noise sources having  $\langle \Gamma_j \rangle=0$ and $\langle \Gamma_i(t) \Gamma_j(s) \rangle = 2 \delta_{ij} \delta(t-s)$.
We have found that, in the flow-free case, the {steady state} probability of finding the dynamical system of Eq.~(\ref{eq:ucnad}) in a specific configuration $\mathbf{x}$ is always proportional to the weight:

\begin{equation} \label{eq:ucnas}
\Omega(\mathbf{x}) = \exp \left[ -\frac{\Phi(\mathbf{x})}{D} -\frac{\tau}{2 D}|\nabla \Phi(\mathbf{x})|^2 \right] 
|| \mathbb{I} + \tau \mathbb{\mathbb{H}} (\mathbf{x}) || 
\end{equation}

\noindent where $|...|$ represents the norm of a vector 
and $||...||$ indicates the absolute value of the determinant of a matrix.
We have demonstrated the validity of Eq.~(\ref{eq:ucnas}) by deriving the 
corresponding Fokker-Planck equation from Eq.~(\ref{eq:ucnad})~\cite{Risken} and solving 
it in the zero-current case by using the Jacobi's formula~\cite{Jacobi} (see Supplemental Material).

\subsection*{One single degree of freedom}

As Eq.~(\ref{eq:ucnas}) is used for specific choices of the potential it reveals several interesting non-equilibrium properties of the active system under study. {We initially focus on a simple one-dimensional case and study GCN-driven 
particles}  when they are subject to a steep repulsive potential of the form 
$\Phi(x)=A x^{-12}$ setting $A=1$. 
Such a potential can be thought as a repulsive obstacle that perturbs the dynamics of the particles~\cite{NumericalBacteria}. 
To verify the quality of the UCNA,
we integrate numerically the stochastic equation of motion (\ref{eq:sys2}) in the presence of such a potential. 
To this aim we have implemented a code for Euler integration of Eq.~(\ref{eq:sys2}), to be executed on GPU {where} the dynamics of many independent particles can be simulated in parallel~\cite{GPU}. 
We consider several different values of $0.1\leq \tau \leq 1 \, \mathrm{s}$  and $0.1\leq D \leq 100 \,   \mathrm{\mu m^2 /s}$,  {ranges that} cover the typical persistence times and diffusivities of colloids in bacterial baths, swimming bacteria such as \textit{E. coli}~\cite{DDMpoon,Centrifugation} and of chemically self-propelled Janus-type particles~\cite{Janus3}. The size of the simulation box is chosen to be $L=20~\mu$m (with periodic boundaries located at $\pm L/2$). 
Fig.~\ref{fig:f1}(a) shows that in equilibrium ($\tau=0$) the probability density decreases rapidly before the core of the repulsive potential is reached, whereas in the GCN the distribution peaks substantially in a region where the potential is very high before vanishing at the core. Note that the specific choice of the constant $A=1$ defines the size of the repulsive ``wall'' created by the external potential. The thickness of this impenetrable region is about $2 \, \mathrm{\mu m}$ and it depends very little on the values of $D$ and $\tau$ considered since the potential is steeply repulsive. 
In the GCN-driven system the exact probability distribution is unknown but it can 
be approximated by Eq.~(\ref{eq:ucnas}) that reduces, for a single
degree of freedom, to the known form~\cite{HanggiRev}:

\begin{equation} \label{eq:u1d}
\Omega(x) = 
\exp \left[-\frac{\Phi(x)}{D}-\frac{\tau }{2 D}| \Phi '(x)|^2 \right] |1+\tau \Phi ''(x)|
\end{equation}

\noindent where the prime indicates differentiation with respect to $x$. 
{The approximate probability, obtained by normalizing Eq.~(\ref{eq:u1d}) 
$P(x)=\Omega(x) / \int dx \Omega(x)$,  is plotted} in Fig.~\ref{fig:f1}(a) 
as a dashed line and it is found to {reproduce} well the numerical distribution at two well 
separated values of $D$ and $\tau$. 
{Knowing the probability we can also compute all the average quantities of interest, such as
 the average value of the modulus of the velocity
$\langle |\Phi '| \rangle = \int dx P(x) |\Phi'(x)|$. The theoretical (approximated) 
$\langle |\Phi '| \rangle$ is compared with the numerical value in Fig.~\ref{fig:f1}(b)
where one sees that the $\langle |\Phi '| \rangle$ UCNA prediction  is very close to the numerically  result at all values of $D$ and $\tau$ here investigated. }
The behaviour of the $P(x)$ and of $\langle |\Phi ' | \rangle$ can be qualitatively understood by considering the strong peaking of the $\Omega(x)$ close to the repulsive barrier. When the potential is very steep, as in the case of $\Phi = x^{-12}$, the maxima of $\Omega(x)$ are found at $x=x^\ast$ where $|\Phi'(x^\ast)| \approx \sqrt{D/\tau}$. It is clear that the probability peaks where the external potential balances the root mean-squared velocity of the particle induced by GCN $\sqrt{\langle \eta^2 \rangle}=\sqrt{D/\tau}$. In this hard-wall limit is possible to approximate $\Omega(x)$ in the neighborhood of $x^\ast$ by a {strongly peaked function $k(x)$ whose integral is $\approx \sqrt{D \tau}$, as found by a saddle-point approximation, while far away from $x^\ast$ Eq.~(\ref{eq:u1d}) reduces to unity (see Supplemental Material) and we can write: $\Omega(x) \approx k(x-x^\ast)+k(x+x^\ast)+1$.}
Note that  $\sqrt{D \tau}$ corresponds to the typical correlation length of the active motion.
{Integrating from $-L/2$ to $L/2$ we find the average velocity $ \langle |\Phi ' | \rangle = 2 D/(L^\ast+2\sqrt{D \tau}) $, where $L^\ast = L-2 x^\ast$ is the overall length available to the particles
and report this result as a dashed line in Fig.~\ref{fig:f1}(b) where it captures the trend of $\langle |\Phi ' | \rangle$ obtained in simulations with the potential }
$\Phi = x^{-12}$. By considering the force exerted by the particles located only on the right-hand side of the potential ($x>0$) we find the force $f_{x>0}$:
$\langle f_{x>0} \rangle = {\mu^{-1}}D/(L^\ast + 2\sqrt{D \tau})$,
{which corresponds to an equation} of state being $f_{x>0}$ the 1-dimensional pressure that the particles exert on the ``wall" represented by the external potential. 
Note that if we consider $N$ independent particles, in the limit $\tau \rightarrow 0$, and set $D=\mu \, k_B T$ we arrive to the ideal gas law in 1\textit{d}: $\langle f_{x>0} \rangle = N k_B T/L^\ast$ and that this equilibrium value constitute an upper bound for the pressure of the active system (see dashed-dotted line in Fig.~\ref{fig:f1}(b)). Interestingly these results can be derived exactly for the RnT model, in particular the stationary probability distribution of the RnT model in presence of two hard walls is composed by two Dirac deltas plus a constant and the expression of $\langle |\Phi'| \rangle$ has the same form of the one found in the UCNA (see Supplemental Material).

\subsection*{Two interacting particles in one dimension.}
\begin{figure}[ht]
\centering
\includegraphics[width=.5\textwidth]{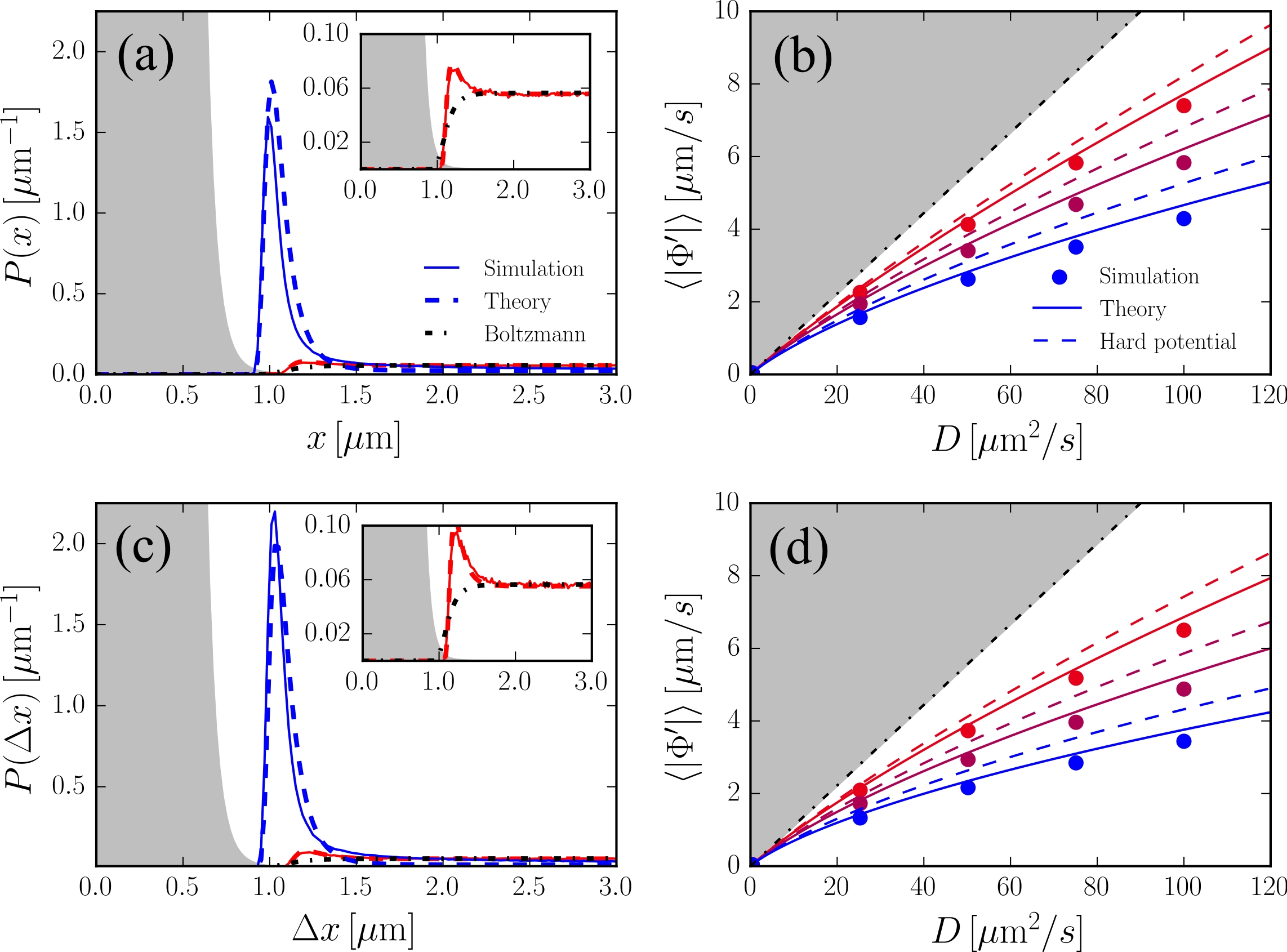}
\caption{\textbf{(a)}
Probability density function of the position of a single GCN-driven particle in presence of the external potential $\Phi=x^{-12}$ (shaded area). 
Full lines: simulations, dashed lines: theory, dashed-dotted line: Boltzmann distribution. The curve with the higher peak corresponds to $D=100 \, \mathrm{\mu m^2/s}$, $\tau=1 \, \mathrm{s}$ and the one with the lower peak to $D=0.4 \, \mathrm{\mu m^2/s}$, $\tau=0.1 \, \mathrm{s}$ (zoomed in the inset)
\textbf{(b)} 
Average value of $|\Phi'|$ as a function of $D$ for three different values of $\tau=0.1,0.325,1$ s from top to bottom respectively. Points: simulations, full lines: theory, dashed lines: theory in the limit of a hard potential, dashed dotted line: white noise case with a hard potential. 
\textbf{(c)}
Probability density function of the distance between two GCN-driven particles 
interacting via the potential $\Phi={\Delta x}^{-12}$, same legend as Fig.~(a).
\textbf{(d)}
Average value of $|\Phi'|$ as a function of $D$ for three different values of 
$\tau=0.1,0.325,1$ s from top to bottom respectively, same legend as as Fig.~(b).}
\label{fig:f1}
\end{figure}

Up to this point we have derived results from the known formula (\ref{eq:u1d}) considering only  one degree of freedom. We now use Eq.~(\ref{eq:ucnad}) to characterize the steady state properties of two interacting GCN-driven particles moving in $1d$. We consider the positions $x_1$ and $x_2$ of two particles interacting via a pair potential that is a function of the distance 
$\Delta x = |x_1-x_2|$ between the particles: $\Phi (x_1,x_2) = \Phi(\Delta x)$. We further assume that the particles are free to move in a 1$d$ space of extension $L$ with periodic boundaries located at $\pm L/2$. In this case Eq.~(\ref{eq:ucnad}) can be used to compute the probability of finding the two particles separated by $\Delta x$:

\begin{equation} \label{eq:u2p}
\Omega(\Delta x) = 
\exp \left[-\frac{\Phi(\Delta x)}{D}-
\frac{2 \tau}{2 D} |\Phi '(\Delta x)|^2 \right] 
|1+2 \tau \Phi''(\Delta x)|
\end{equation}

\noindent which is identical { to Eq.~(\ref{eq:u1d}) with  $\tau$ replaced by $2 \tau$. This is at variance} with equilibrium statistical mechanics in which the probability of finding two particles at a given distance does not vary if one of the two particles is pinned at some fixed position~\cite{RandomPinning1,RandomPinning2,RandomPinning3}. 
To be more specific let us consider a repulsive potential of the form $\Phi(x_1,x_2)=(x_1-x_2)^{-12}$. In this case again the probability 
is well described by the MUCNASP as shown in Fig.~\ref{fig:f1}(b). Similarly the deterministic velocity component experienced by one particle $\langle |\Phi'| \rangle$ resulting from simulations is well approximated by the theory (see Fig.~\ref{fig:f1}(c)). Again, in the limit of an infinitely steep potential (see Supplemental Material), we find that  the area of the peak of $\Omega(\Delta x)$ is approximated by $\sqrt{2 D  \tau}$ and $ \langle |\Phi ' | \rangle = 2 D/(L^\ast+2\sqrt{2 D  \tau})$ which is plotted in Fig.~\ref{fig:f1}(c) as a dashed line.
This can be physically interpreted as follows: when both particles are free to move they can be more often found in contact since they move {coherently} in the same direction, this happens without the particles pushing onto each other, yielding a lower value of the average interaction force. It is important to note that these theoretical results cannot 
be derived {analytically in the RnT model since the coupled dynamics of more particles} makes the problem far too complicated. Nevertheless we have found that the MUCNASP produces results for the average $\langle | \Phi' |\rangle$ that are very similar those found numerically for the RnT model despite the stationary probability density has a very different form (see Supplemental Material). This suggests that the MUCNASP can be used as a convenient approximation also for calculating the averages in RnT dynamics at least in the case of steeply repulsive potentials. Note also that such a scenario is in agreement with the findings of Ref.~\cite{ActiveDepletion} where it was demonstrated, by combining experiments and simulations, that two colloids suspended in a bacterial bath tend to stay in contact because of the colored-noise forces induced by swimming bacteria.

\begin{figure}[ht]
\centering
\includegraphics[width=.5\textwidth]{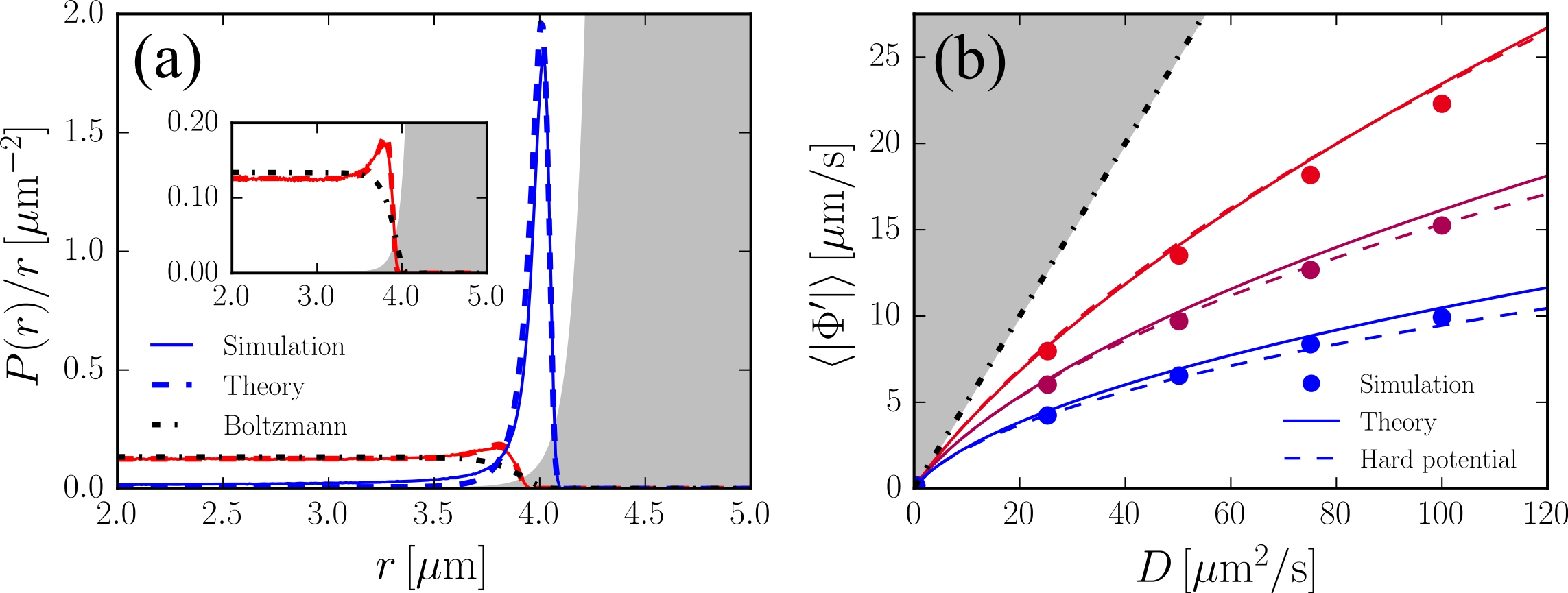}
\caption{\textbf{(a)}
Probability density function of the radial distance of one GCN-driven particle in presence of the spherically symmetric external potential $\Phi=(r-R)^{-12}$ in $2d$ (shaded area). 
Full lines: simulations, dashed lines: theory, dashed-dotted line: Boltzmann distribution. The  curve with the higher peak corresponds to $D=100 \, \mathrm{\mu m^2/s}$, $\tau=1 \, \mathrm{s}$ and the  one with the lower peak to $D=0.4 \, \mathrm{\mu m^2/s}$, $\tau=0.1 \, \mathrm{s}$ (zoomed in the inset)
\textbf{(b)} 
Average value of $|\Phi'|$ as a function of $D$ for three different values of $\tau=0.1,0.325,1$ s from top to bottom respectively. Points: simulations, full lines: theory, dashed lines: theory in the limit of a hard potential, dashed dotted line: white noise case with a hard potential.}
\label{fig:f2}
\end{figure}

\subsection*{Radially symmetric potentials.}
When the $d$-dimensional potential is spherically symmetric, i.e. 
$\Phi(\mathbf{x})=\Phi(r)$, Eq.~(\ref{eq:ucnas}) simplifies to

\begin{eqnarray}
\Omega(r)  = & &  
\Theta \, \exp \left[-\frac{\Phi(r)}{D}-
\frac{\tau}{2 D} |\Phi '(r)|^2 \right]
|1+ \tau \Phi''(r)| \nonumber\\
& & |r+ \tau \Phi'(r)|^{d-1} \label{eq:rad}
\end{eqnarray}

\noindent where $r^2=\sum_{i=1}^d {x_i}^2$ and $\Theta$ is the d-dimensional  solid angle.
 Note that the Boltzmann distribution, obtained by setting $\tau=0$ in 
Eq.~(\ref{eq:rad}), depends on the dimensionality only via the trivial  term $r^{d-1}$
while in the GCN-case this dependence is more complicated. {To understand this issue, we consider GCN-driven particles in $d=2$ in the presence of a circular repulsive potential of radius $R$ of the form  $\Phi(r)=(r-R)^{-12}$ where $r=\sqrt{x^2+y^2}$ and $R=5 \, \mathrm{\mu m}$.  Simulation results
show that  particles accumulate near the ring at $r=R$} and the theoretical probability reproduces well this behaviour (see Fig.~\ref{fig:f2}(a)).
From Eq.~(\ref{eq:rad}) we can compute the averages of interest as the radial component of the  velocity field $\langle | \Phi' | \rangle$ and   compare it with simulation results in Fig.~\ref{fig:f2}(d) showing a good agreement.
In the limit of an infinitely steep potential we have that $\Omega(r)$   strongly  peaks  where 
$|\Phi '| \approx \sqrt{D / \tau}$ and reduces to unity elsewhere. The area of the peak can be approximated by $2 \pi \sqrt{D \tau} |R^\ast+\sqrt{D \tau}|$ (see Supplemental Material), where $ R^\ast = R-r^\ast $  
is the radial coordinate of the peak with $ r^\ast \approx 1$ in the $D$-$\tau$ range considered. For the average radial velocity component we get 
$\langle | \Phi '| \rangle = 
2 D (R^\ast+\sqrt{D \tau})/({R^\ast}^2+2 D \tau +2 R^\ast \sqrt{D \tau})$
which is plotted as dashed lines in Fig.~\ref{fig:f2}(d) and follows nicely the trend displayed by the numerical data. This is practically a colored-noise version of the ideal gas law in a circular container. This is clear if we set $\tau = 0$ to obtain $\langle | \Phi' | \rangle = 2 D/R^\ast$ which is proportional to the average radial force $\langle f \rangle = 2 D/(R^\ast \mu)$. Dividing this by the 2$d$ ``surface" $2 \pi R^\ast $ we get the ideal gas pressure 
$p=N \langle f \rangle /(2 \pi R^\ast) = N k_B T /(\pi {R^\ast}^2)$ for $N$ independent particles and $D=\mu k_B T$. The unperturbed dynamics of the RnT model in $2d$ is well understood~\cite{RnT2d}, while the problem of a 2$d$ symmetric potential is not tractable analytically. However we have found that the simulation results for the $\langle |\Phi'| \rangle$ in RnT model are very similar to those produced by the MUCNASP (see Supplemental Material) suggesting that MUCNASP could actually describe the behaviour of a wider class of active particles. An experimental situation, that is close to the spherically symmetric case treated here, has been recently studied~\cite{Drop}. In these experiments it was observed a marked accumulation of swimming bacteria at the border of spherical liquid droplets. In this kind of experiment it would be interesting to check whether, in the dilute regime, the number of bacteria found in contact with the border  scales with the droplet radius and the characteristic run length as predicted by the MUCNASP (Eq.~(\ref{eq:rad})) in the spherical case.

\section*{Discussion}

By using the unified colored noise approximation, we have derived the explicit formula for the non-equilibrium stationary probability (MUCNASP, Eq.~(\ref{eq:ucnas})) of a system composed by an arbitrary number of degrees of freedom subject to GCN. 
We have focused onto the case of steep repulsive potentials where the probability distribution of one single active particle tends to concentrate on the repulsive part of the potential oppositely to the case of a Brownian particle. 
Moreover we have verified that, as predicted by the MUCNASP, two active particles interacting repulsively behave differently with respect to equilibrium and are found in contact more often than if one particle is fixed. Finally the MUCNASP predicts that, when active particles are confined inside a repulsive ring-shaped barrier, the probability peaks on the boundary and the area of this peak increases with increasing radius and with increasing persistence length. 
Surprisingly the results obtained by the MUCNASP are very close to those obtained for RnT particles for which an analytical solution is not available.
As discussed before~\cite{HanggiRev,HanggiUcna,UcnaD}, the UCNA is accurate in those 
portions of phase space where the all the eigenvalues of the Hessian matrix are positive.
This restriction defines where Eq.~(\ref{eq:ucnas}) can be used as a valid approximation for the probability of GCN-driven system. Moreover it appears difficult to derive an explicitly formula for probability including also Brownian fluctuations. Nevertheless, to our knowledge, the MUCNASP is the only available explicit probability formula accounting for multiple active degrees of freedom and, provided that all eigenvalues are positive, becomes exact both in the limit of $\tau \to 0$ and $\tau \to \infty$. This makes the MUCNASP a valuable schematic model for tackling the many-body problem in active matter. For example it would be interesting to check whether, at the mean-field level, the MUCNASP can predict a motility-induced phase separation~\cite{CatesPRL,Pago,BerthierMI} or the colored-noise induced shift in the synchronization threshold of the noisy Kuramoto model~\cite{Kuramoto}.


\section*{Acknowledgements}

The research leading to these results has received funding from
the European Research Council under the European Union's
Seventh Framework Programme (FP7/2007-2013)/ERC grant
agreement no. 307940. 
NG acknowledges support from MIUR
(``Futuro  in  Ricerca"  ANISOFT/RBFR125H0M).
We also acknowledge NVIDIA for hardware
donation.



\section*{Additional information}

\textbf{Competing financial interests} The author(s) declare no competing financial interests.

\end{document}